\documentclass{elsart}
\usepackage{epsfig}
\usepackage{amssymb}
\usepackage{amsmath}
\usepackage{subfigure}
\usepackage{float}

\begin{document}
\begin{frontmatter}
\vspace*{-1.5cm}

\title{Spectator Detection for the Measurement of Proton-Neutron Interactions
at ANKE}

\author[FZJ,FZR]{I.~Lehmann\thanksref{label1}},
\thanks[label1]{Current address: Department of Radiation Sciences,
University of Uppsala,\\ Box 535, S-75121 Uppsala, Sweden}
\author[PNPI]{S.~Barsov},
\author[FZJ]{R.~Schleichert}\ead{r.schleichert@fz-juelich.de},
\author[UCL]{C.~Wilkin},
\author[ZEL]{M.~Drochner},
\author[FZJ]{M.~Hartmann},
\author[FZJ]{V.~Hejny},
\author[Dubna]{S.~Merzliakov},
\author[PNPI]{S.~Mikirtychiants},
\author[FZJ]{A.~Mussgiller},
\author[FZJ]{D.~Proti\'c},
\author[FZJ]{H.~Str\"oher},
\author[MSU,FZR]{S.~Trusov},
\author[ZEL]{P.~W\"ustner}

\address[FZJ]{Institut f\"ur Kernphysik, Forschungszentrum J\"ulich,
D-52425 J\"ulich, Germany}
\address[FZR]{Institut f\"ur Kern und Hadronenphysik,
Forschungszentrum Rossendorf,\\ D-01314 Dresden, Germany}
\address[PNPI]{High Energy Physics Department, Petersburg Nuclear
Physics Institute,\\ 188350 Gatchina, Russia}%
\address[UCL]{Physics \& Astronomy Department, UCL, Gower Street, London
WC1E 6BT, UK}
\address[ZEL]{Zentralinstitut f\"ur Elektronik, Forschungszentrum J\"ulich,
D-52425 J\"ulich, Germany}
\address[Dubna]{Laboratory of Nuclear Problems, Joint Institute for Nuclear
Research, Dubna, 141980 Dubna, Russia}
\address[MSU]{Skobeltsin Institute for Nuclear Physics of 
M.~V.~Lomonosov Moscow State University, Vorobjovy Gory,
119899 Moscow, Russia}

\begin{abstract}

  A telescope of three silicon detectors has been installed close to the
  internal target position of the ANKE spectrometer, which is situated inside
  the ultra-high vacuum of the COSY-J\"ulich light-ion storage ring. The
  detection and identification of slow protons and deuterons emerging from a
  deuterium cluster-jet target thus becomes feasible. A good measurement of
  the energy and angle of such a \textit{spectator} proton ($p_\mathrm{sp}$)
  allows one to identify a reaction as having taken place on the neutron in
  the target and then to determine the kinematical variables of the
  ion-neutron system on an event-by-event basis over
  a range of c.m.\ energies.

The system has been successfully tested under laboratory conditions. 
By measuring the spectator proton in the $pd\to p_\mathrm{sp}d\pi^0$ 
reaction in coincidence with a fast deuteron in the ANKE Forward Detector, 
values of the $pn\to d\pi^0$ total cross section have been deduced.
Further applications of the telescope include the determination of 
the luminosity and beam polarisation which are required for several 
experiments.

\end{abstract}


\end{frontmatter}

\section{Introduction}

In any thorough investigation of meson production or other
reactions resulting from nucleon-nucleon collisions, it is
important to have proton-neutron data as well as proton-proton.
Medium energy neutron beams, even when produced from stripped
deuterons, have a significant momentum spread. Unless the $np$
centre-of-mass energy is determined by other means, such as by
measuring all final-state particles, these neutron beams are not
suitable for measuring cross sections that vary fast with energy,
as is the case for meson production near threshold. The
alternative, of using the neutron inside deuterium as a target for
a proton beam, faces similar problems due to the Fermi motion of
the neutron in the target. Despite the typical neutron momentum in
the deuteron being only around 60~MeV/c, this spreads the c.m.\
energy by approximately 100~MeV, depending upon the mass of the
meson being produced. A precise experiment therefore requires all
final-state particles to be measured also in this approach.

The suggestion of using the deuteron as a substitute for a neutron
target is very plausible because the average proton-neutron
separation in the nucleus is about 4~fm, which is large compared
to the typical range of forces between elementary particles. To a
good first approximation then, an incident particle will interact
with either a proton or neutron target, leaving the other nucleon
in the deuteron as a \textit{spectator}, moving with the momentum
that it had before the collision. In the thick targets used in
bubble chamber or electronic measurements with external beams, one
rarely detects such a spectator particle, which has an energy of
only a few MeV. This situation has been radically changed with the
advent of medium energy proton storage rings where one can work
with thin windowless targets. It then becomes feasible to measure
the proton spectators in solid state counters placed in the vacuum
of the target chamber inside such a ring. Knowing the beam energy
and the spectator momentum allows one to reconstruct the c.m.\
energy of the proton-neutron system without measuring all the
particles produced in this reaction. In this way it is possible to
exploit the intrinsic target momenta, measuring a cross section
over a range of well-defined c.m.\ energies while working at a
fixed beam momentum.  Furthermore, merely identifying a spectator
proton $p_\mathrm{sp}$ is an indication that the production
reaction has taken place on the neutron and not on the proton, thus
simplifying significantly the subsequent data analysis.

These ideas have all been successfully tested in an initial
experiment at the CELSIUS ring, where the spectator protons were
measured in a set of simple silicon detectors~\cite{Tord}. All
final particles in the $pd\to p_\mathrm{sp}d\pi^0$ reaction were
detected, with the two photons from the $\pi^0$ decay being used
as the trigger. Using a circulating beam whose energy was fixed at
320~MeV, the authors could deduce within the spectator model the
variation of the $pn\to d\pi^0$ total cross section at five c.m.\
energies near threshold and this agreed with the results from
standard data compilations. On the other hand, the lack of an
independent luminosity measurement meant that it was not possible
to determine absolute cross sections. Since there is no room for
these spectator counters in the new WASA target chamber at
CELSIUS, they are now being used at COSY to study the $pn\to
pn\eta'$ reaction~\cite{COSY11}.

At the COSY-ANKE facility of the Forschungszentrum J\"ulich we
have extended the CELSIUS technique by constructing a telescope of
three silicon detectors, the details of which are given in
section~2. The self-triggering capabilities of the telescope
allow efficient data taking at high particle fluxes. As discussed
in the following section, the set-up yields good measurements of
both the energies and angles of slow protons and deuterons
emerging from a deuterium target. Using the $E$-$\Delta E$
technique the system permits particle identification and, in
particular, the separation of recoil protons from deuterons over
significant energy ranges. One new feature of this set-up is its
capability of working down to proton momenta of 70$\,$MeV/c. Not
only are the achievable statistics significantly increased, as
compared to experiments with higher threshold energies, but also
the interpretation of the data in terms of its model dependence
becomes substantially simplified.

The technique is put to the test in section~4, where the
measurement of the $pd\to p_\mathrm{sp}d \pi^0$ reaction away from
threshold is described. By using a deuterium cluster-jet target
and detecting the fast deuteron in the forward system of the ANKE
magnetic spectrometer, the $\pi^0$ could be identified as a peak
in the $p_\mathrm{sp}d$ missing mass. The $pn\to d\pi^0$ total
cross section thus extracted agrees with the results deduced from
previous experiments.

In subsequent experiments the spectator telescope has been used to
measure the total cross sections for the $pn\to
d\omega$~\cite{omega,PhDIL} and $pn\to d\eta$~\cite{eta} reactions
near threshold.  At the high beam momenta required for $\omega$
production, the recoil deuterons from elastic proton-deuteron
scattering can also be detected in the telescope and used to
determine the luminosity by comparing with known $pd\to pd$
differential cross sections. Another application of the set-up is
the measurement of the beam polarisation, which is vitally
important for the polarised experimental programme at ANKE. These
measurements, and the outlook for further work, are outlined in
our conclusions of the final section.\clearpage

\section{Silicon Telescope} \label{telescope}

\begin{figure}[H]
\begin{center}
\epsfig{file=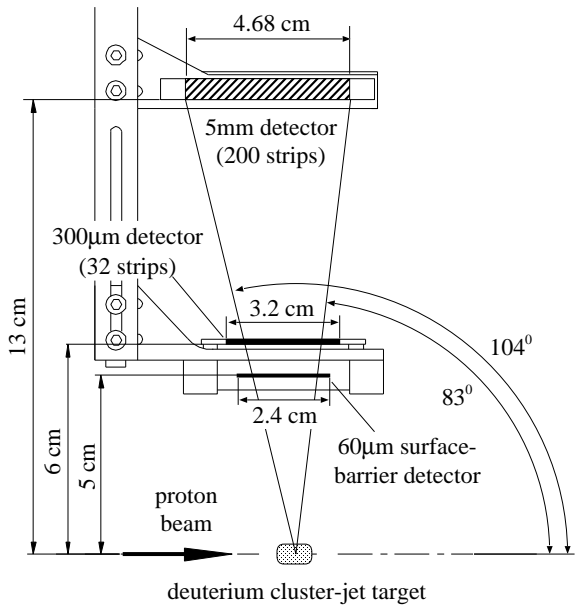,height=8cm}
\caption{Sketch of top view of the silicon telescope inside the ANKE
target chamber showing the COSY beam, the cluster target and the
telescope structure of three silicon detectors.}
\label{f:SP_setup}
\end{center}
\end{figure}

The ANKE facility is a magnetic spectrometer and detection system
placed at an internal target position of the COSY light-ion
storage ring~\cite{ANKE}. In order to extend its capabilities
through the detection of recoil particles with very low momenta, a
telescope consisting of three layers of silicon
detectors has been chosen, as sketched in top view in
Fig.~\ref{f:SP_setup}.  This enables tracking, energy
determination, and particle identification to be carried out
within the constraints imposed by the restricted space inside the
ANKE vacuum chamber.

In order to identify a particle \textit{via} the $E$-$\Delta E$
method and to determine its kinetic energy, it has to be stopped
after traversing a preceding detector and depositing part of its
energy therein. Now the Fermi momentum of a spectator proton in
the deuteron is on average about 60$\,$MeV/c ($T_p\approx
1.9\,$MeV) and only about 5\% of protons have momenta above
250$\,$MeV/c. Moreover, at the higher momenta one has problems in
being sure that the recoil proton is indeed a spectator rather
than an active participant in a reaction.

A 5$\,$mm thick lithium-drifted strip detector~\cite{Protic} stops
protons with kinetic energies up to 31$\,$MeV, \textit{i.e.}
momenta up to $250\,$MeV/c. Taken in combination with a
300$\,\mu$m thick detector, protons with $T_p \ge 6.7\,$MeV could
in principle be used. However, as will be seen in the following
section, detection thresholds and angular straggling
considerations increase this limit to $8\,$MeV. In order to extend
this range down to kinetic energies of 2.5$\,$MeV, an initial
layer consisting of a 60$\,\mu$m thick surface-barrier detector
was also installed. This choice combines a low threshold for
protons, with energy deposits sufficient to distinguish protons
from deuterons. Though it would have been very helpful to reduce
the energy threshold for spectator protons by using, say, an
18$\,\mu$m thick surface-barrier detector, it was found in a
separate test measurement~\cite{DiplIL} that the energy resolution
\textit{versus} band separation in this case would not suffice to
distinguish between protons and deuterons. The details of the
layers finally chosen are given in Table~\ref{t:si_detectors}.

\begin{table}[H]

\caption{Principal properties of the detectors used in the
spectator telescope. The first layer is circular and not
segmented. For the others, the number of strips and their pitch,
is given. The energy resolution of the detectors was deduced from
measurements with an $\alpha$-particle source, and the full-width
at half maximum (FWHM) of the peaks, which characterises the noise level, 
is given.}

\label{t:si_detectors}
\vspace{3mm}
\begin{center}
\begin{tabular}[c]{l|l l l}
\hline
 & $1^\mathrm{st}$ layer & $2^\mathrm{nd}$ layer & $3^\mathrm{rd}$ layer \\
\hline
Silicon detector type      & Surface barrier & Implanted       & Lithium-drifted \\
Sensitive thickness        & $60.9\,\mu$m  & $306\,\mu$m       & $5.1\,$mm\\
Entrance\ window [Si-eqv.] & 0.08$\,\mu$m  & $\leq 1.5\,\mu$m  & $\le 1\,\mu$m \\
Exit window [Si-eqv.]      & 0.23$\,\mu$m  & $\leq 1.5\,\mu$m  & $\le 1\,$mm \\
Active area                & $450\,$mm$^2$ & 32$\, \times \,15\,$mm$^2$ &
47$\, \times \,23\,$mm$^2$ \\
Segmentation        & 1           & 32             & 200\\
Pitch               & ---            & $1\,$mm             & $235\,\mu$m \\
Noise               & 100$\,$keV & 70$\,$keV & 80$\,$keV \\
\hline
\end{tabular}
\end{center}
\end{table}

With respect to the incident beam direction, angles from the
deuterium cluster-jet target between 83 and 104$^\circ$ are
covered in the horizontal plane, whereas in the vertical direction
limits of $\pm 10^\circ$ and $\pm 7^\circ$ are set by the second
and third layers respectively. Having some coverage in the forward
hemisphere is important because deuterons from elastic
proton-deuteron scattering, used for luminosity determination at
high energies as well as for the alignment, cannot go backwards in
the laboratory frame.\\

Simple kinematic arguments show that, for counters placed close to
$90^{\circ}$, both the c.m.\ energy $W_{pn}$ of the proton-neutron
system and the typical missing-mass determinations depend even
more sensitively upon the polar angle of the spectator proton than
on its energy. To handle this, the second and third layers of the
telescope are composed of strips arranged perpendicularly to the
beam. The surface-barrier detector and all 32 strips of the
300$\,\mu$m thick detector are read out individually using
analogue electronics.
Preamplifiers with a gain of 55$\,$mV/MeV are placed outside the 
vacuum about 10$\,$cm from the detectors. The shaper amplifiers, 
with shaping constants of 1$\,\mu$s differentiating and 1$\,\mu$s integrating,
are placed 20$\,$m apart.
The latter 32 channels have additionally been equipped with fast shaper amplifiers 
(shaping constants: 1$\,\mu$s differentiating, 10$\,$ns integrating)
and discriminators to allow triggering. 
This feature is essential for efficient data-taking at high particle fluxes 
because the elastic and quasi-free elastic channels cannot be efficiently
suppressed by the on-line trigger from the ANKE forward system alone.
The last layer, however, has been equipped with 4 resistor
chains such that 8 read-out channels are sufficient
to obtain the energy loss and coordinate, using the sum and
difference of amplitudes respectively~\cite{DiplIL,res_chain}. 
Due to the higher dynamic range, the gain of the preamplifiers is only
5$\,$mV/MeV with an unchanged shaping constant of 1$\,\mu$s
of the main amplifieres.
The angular resolution that can be obtained with the chosen arrangement will
be discussed in the next section.

One big advantage of our set-up is that it allows us to determine
the relative positions of the target and detectors far more
accurately than can be done by conventional techniques. This is
because the elastic proton-deuteron kinematics are fully determined by
the measurement of the deuteron kinetic energy in the second and
third layers of the telescope. For this reason an accurate energy
calibration of the system was performed, as described below. With
the achieved precision of 1\% in the particle's kinetic energy,
the telescope could be aligned to within a few tenths of a
millimetre with respect to the target. Moreover, the angles of the ANKE
forward system could also be determined with respect to the target
and beam to about 0.05$^\circ$.  Though this feature is useful
when determining the spectator angles, it is essential for
reliable acceptance calculations.

\begin{figure}[H]
\begin{center}
  \subfigure[Calibration using a pulser and an $\alpha$-particle source]{
    \epsfig{file=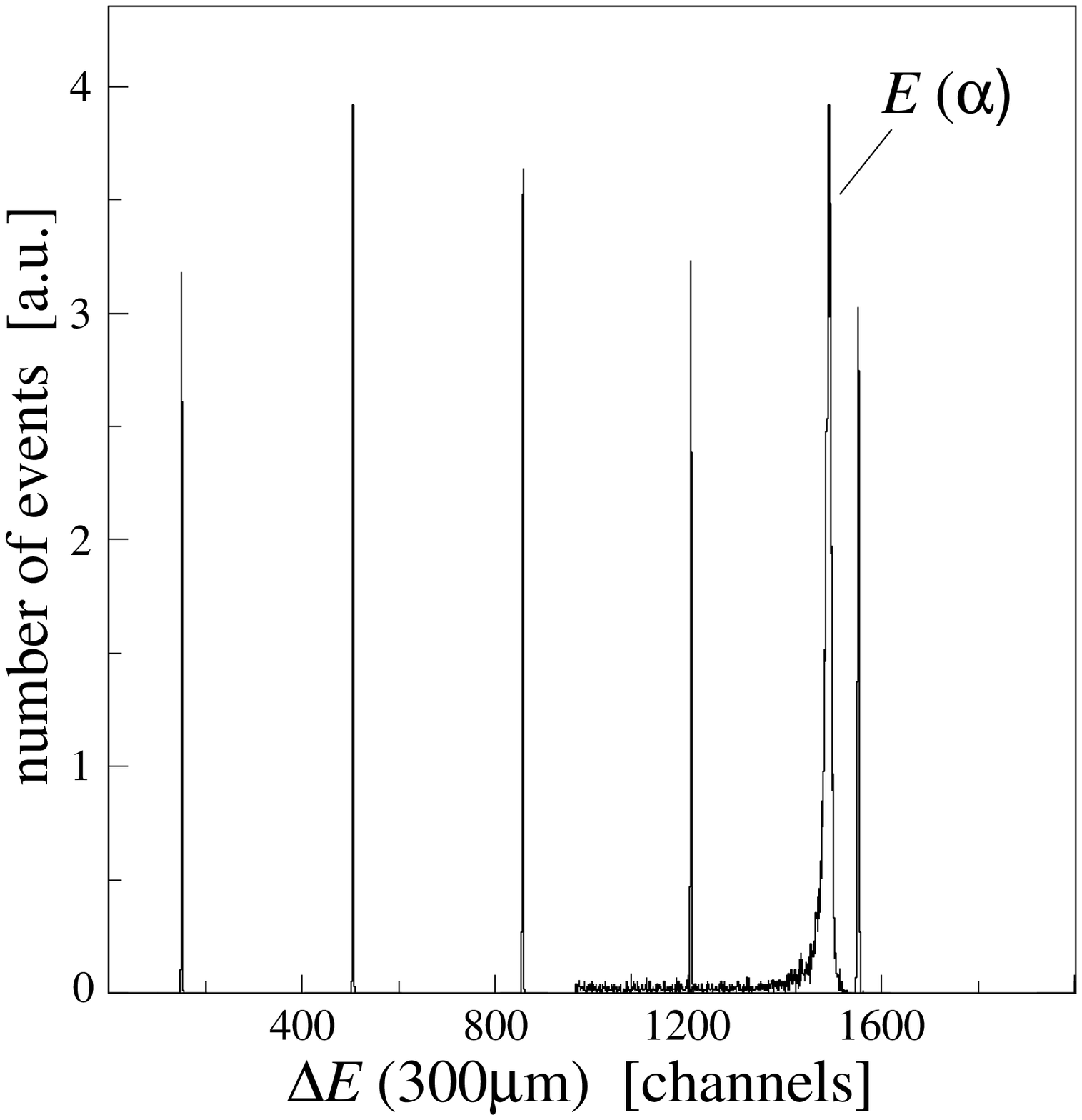,height=6.2cm} \label{f:SP_pulser}}
  \subfigure[On-line calibration check]{
    \epsfig{file=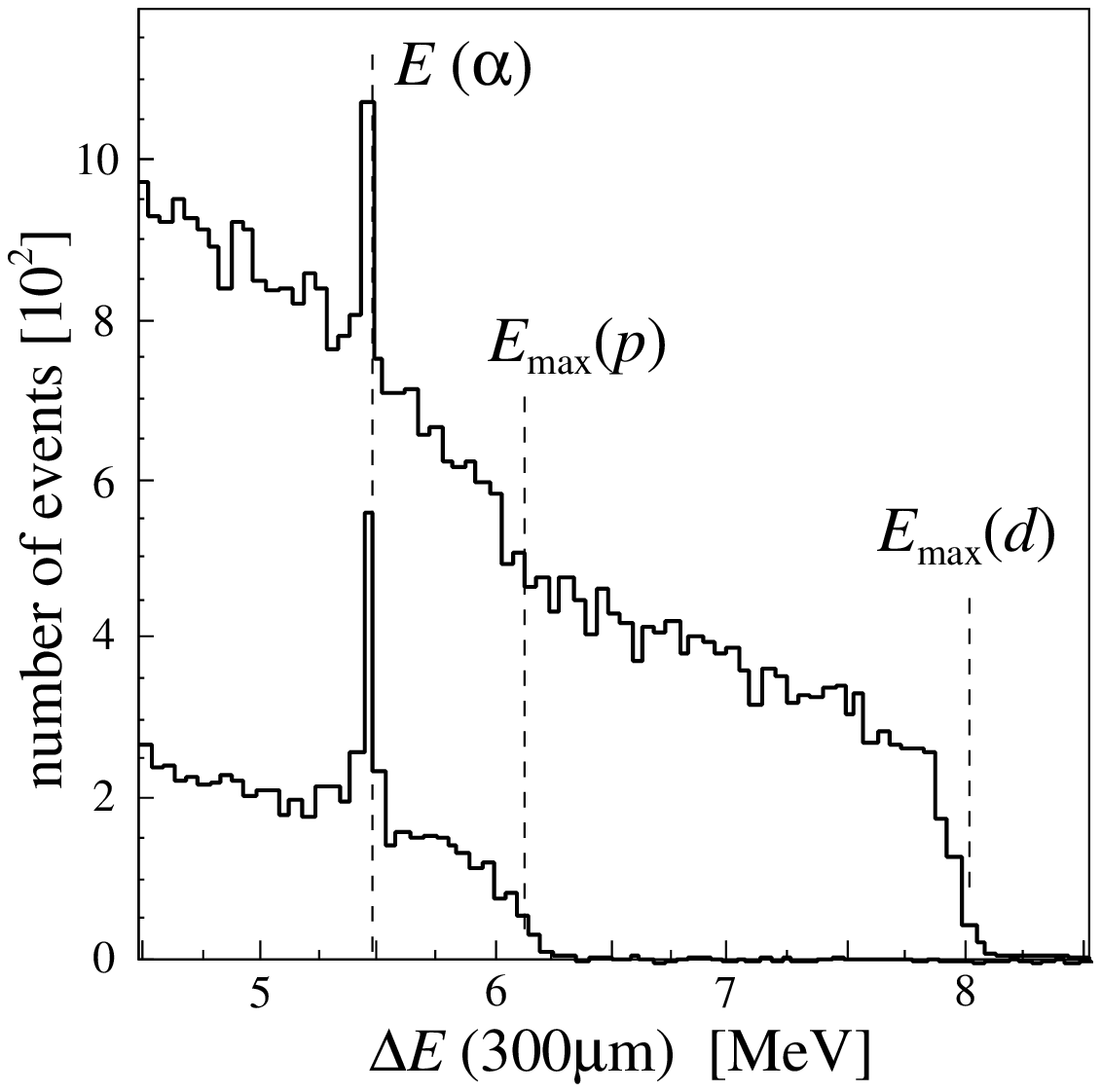,height=6.2cm} \label{f:SPalpha}}

  \caption{Laboratory calibration and on-line check for the second
    telescope layer. (a) The pulser signals and $\alpha$-peak position
    permit an absolute off-line calibration. The calibration at the
    accelerator can be achieved using the calibrated pulser signals
    and cross-checked by the experimental data shown in (b). The upper
    distribution was obtained by selecting forward angles, where
    elastically scattered deuterons are present, whereas only protons
    can contribute in the backward hemisphere shown in the lower
    distribution.  The absolute energy calibration was controlled by
    the well-defined energy loss of stopped $\alpha$-particles coming
    from permanently mounted sources $E ( \alpha )$ (narrow peaks) and
    by the maximum energy loss for protons and deuterons,
    $E_\mathrm{max} (p)$ and $E_\mathrm{max} (d)$ respectively.}

\end{center}
\end{figure}

The amplitudes for each read-out channel were calibrated
individually using $\alpha$-particle sources in combination with
an electronic pulser designed to deliver precise attenuation
factors. Taken together with the known energy deposit from the
$\alpha$-particles, an absolute calibration was then obtained
including all the non-linearities of the system, as shown e.g.\
for the second layer by Fig.~\ref{f:SP_pulser}. The pulser
amplitudes have been calibrated simultaneously, allowing them to
be used to calibrate the complete system at any time after their
installation at COSY. In order to monitor the calibration of the
300$\,\mu$m and 5$\,$mm thick detectors simultaneously with the
measurements, two low-intensity $\alpha$-particle sources (50 and
500$\,$Bq) were permanently mounted in the system. The strengths
were such that, while being negligible compared to the primary
count rate, more than 300 events could be collected per read-out
channel within a typical run-time of two hours. These yielded
clean peaks above the background, as shown in
Fig.~\ref{f:SPalpha}. \clearpage

\section{Spectator Proton Detection} \label{spec}

\begin{figure}[H]
 \begin{center}
  \subfigure[Energy losses in the 60$\,\mu$m {\em versus}
   300$\,\mu$m silicon detectors]
   {\epsfig{file=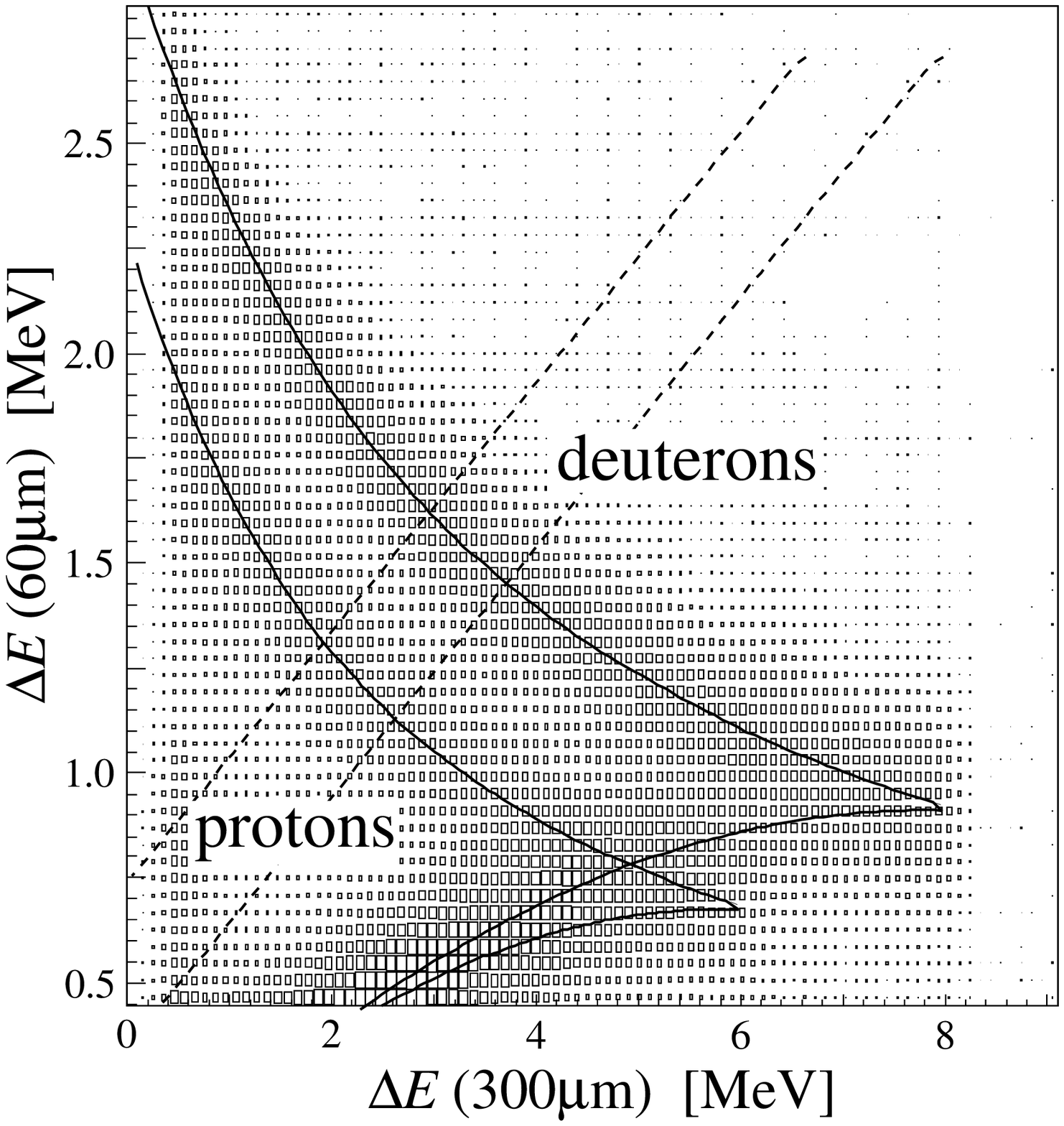,height=6.7cm}
   \label{f:SP_dE/E12}}
  \subfigure[Energy losses in the 300$\,\mu$m {\em versus} 5$\,$mm
   silicon detectors]
   {\epsfig{file=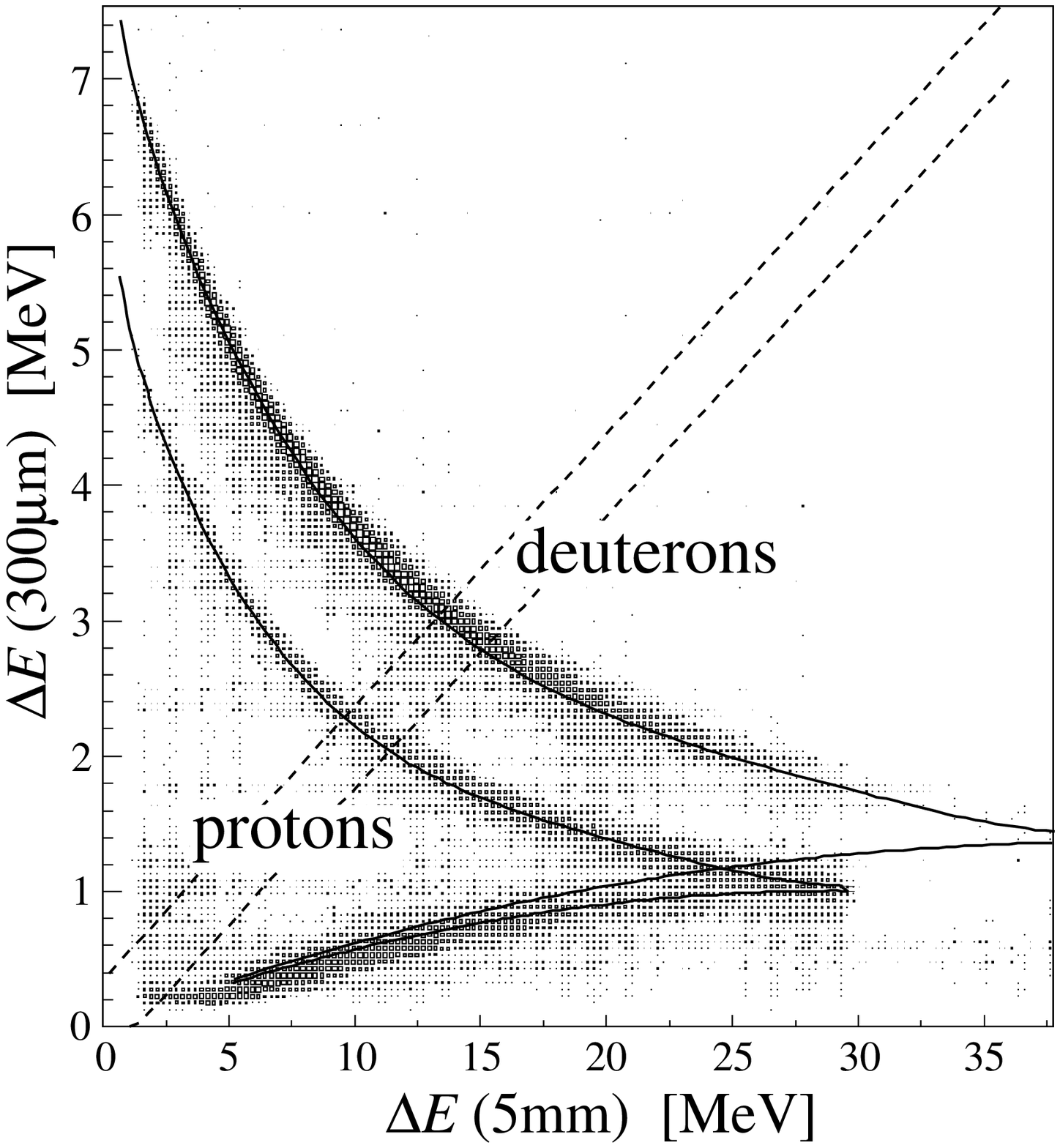,height=6.7cm}
   \label{f:SP_dE/E23}}

  \caption{Particle identification achieved by comparing energy
    deposits in two layers of the telescope. The boxes represent
    experimental data whereas the curves correspond to the predicted
    energy losses of protons and deuterons in silicon~\cite{SRIM}. The
    deuteron band in (b), obtained at $2.8\,$GeV/c, largely disappears
    at low beam momenta due to the restricted geometrical acceptance
    of the system; the recoil deuterons from elastic $pd$ scattering
    then generally have too little energy. The cuts applied to produce
    Figures~\ref{f:SP_sep} are indicated by the dashed lines.}

  \label{f:SP_dE/E}
 \end{center}
\end{figure}

Particle identification is carried out using the $E$-$\Delta E$
technique, as illustrated in Fig.~\ref{f:SP_dE/E}. The agreement
between the experimental data and the predicted proton and
deuteron curves for the two pairs of counters prove that the
systematic error in the absolute kinetic energy determination is
below 1\%.

\begin{figure}[H]
 \begin{center}%
  \subfigure[Combination of $1^\mathrm{st}$ \& $2^\mathrm{nd}$ layers]
    {\epsfig{file=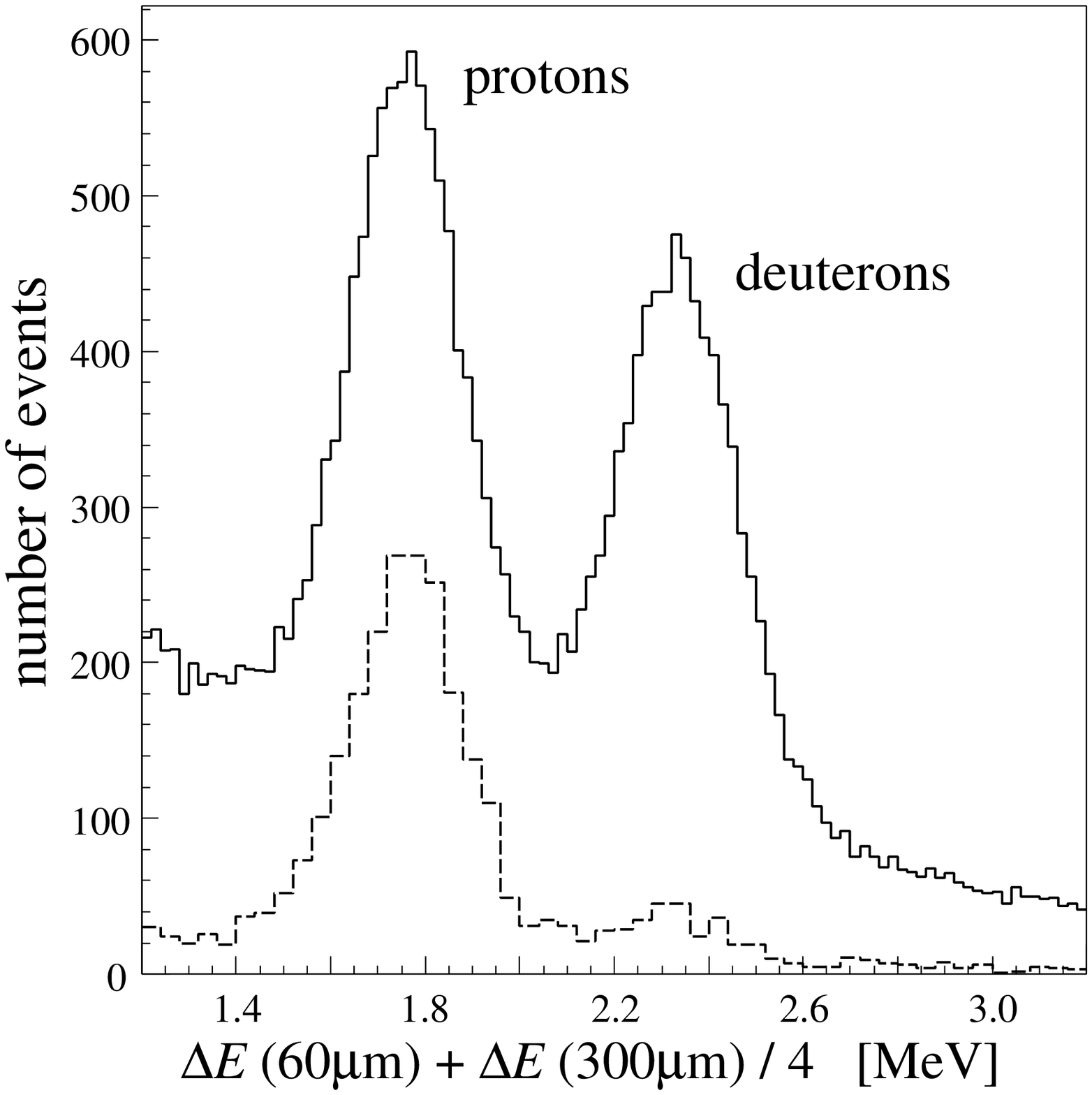,height=6.7cm}}
  \subfigure[Combination of $2^\mathrm{nd}$ \& $3^\mathrm{rd}$ layers]
    {\epsfig{file=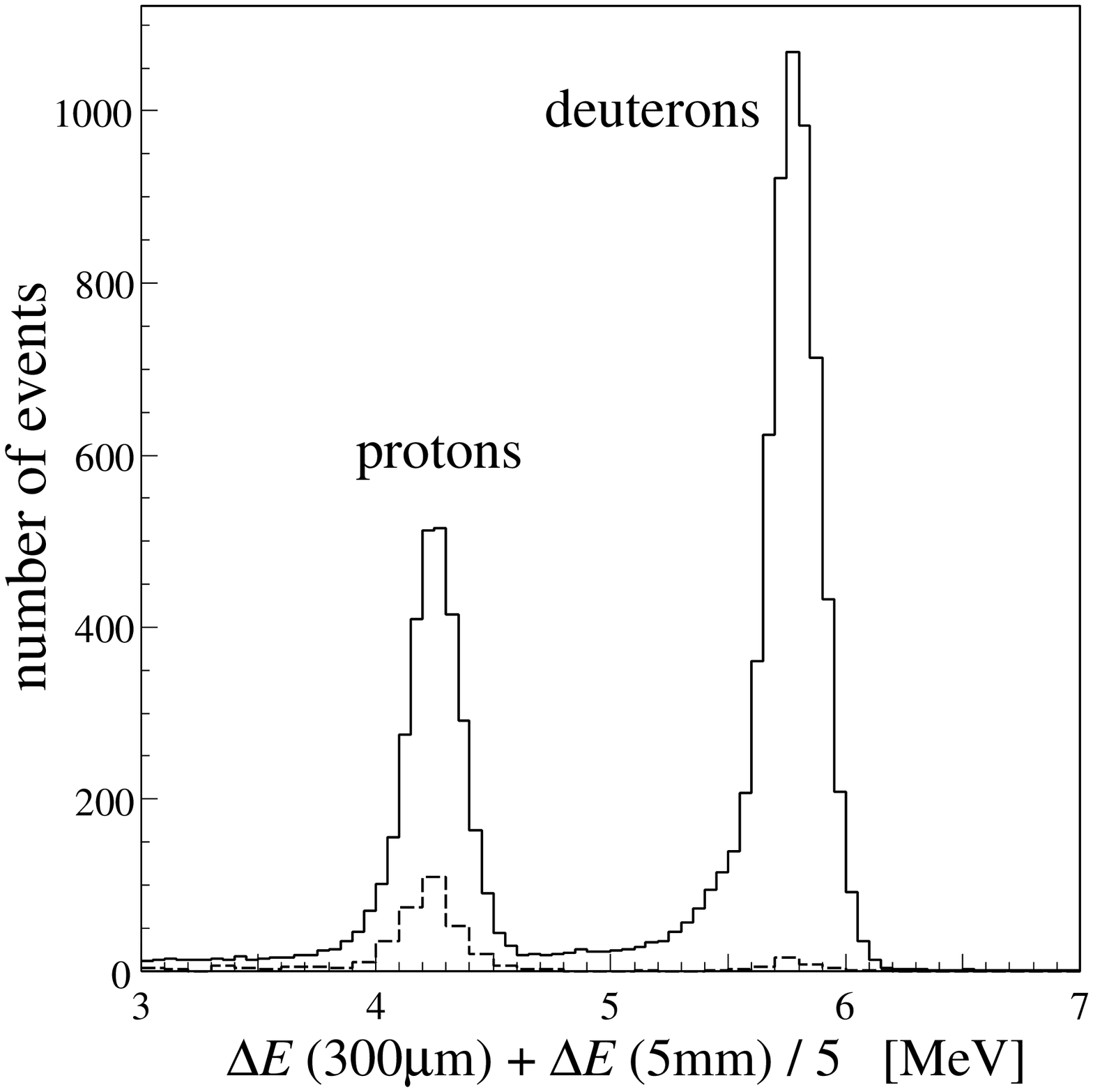,height=6.7cm}}

  \caption{Projections of the slices in the $E$-$\Delta E$
   spectra obtained by applying the cuts indicated by the dashed
   lines in Figs.~\ref{f:SP_dE/E}. These correspond to the combinations
   $\Delta E(60\,\mu\textrm{m})+\Delta E(300\,\mu\textrm{m})/4$ and
   $\Delta E(300\,\mu\textrm{m})+\Delta E(5\,\textrm{mm})/5$ respectively.
   The solid histograms, scaled by factors of 0.2 and 0.05 for (a) and
   (b) respectively, show the separation of the proton
   and deuteron bands in both energy ranges. The
   dashed histograms are obtained after the rejection of elastic events on
   the basis of information obtained from the forward system.}%

  \label{f:SP_sep}
 \end{center}
\end{figure}

To quantify the background arising from possible
misidentification, slices have been cut from Fig.~\ref{f:SP_dE/E}
and the resulting projections plotted in Fig.~\ref{f:SP_sep}. The
effective resolution, including all experimental contributions
such as straggling, detector and electronic response, {\em etc.},
was found to be such that the bands are separated over our full
range by more than two and six times their widths for the low and
high energy particles respectively. From this it is already clear
that, by applying an appropriate cut between the bands, protons
can be well separated from deuterons.  The residual deuteron
background becomes completely insignificant if elastic events are
further suppressed by information from the ANKE forward system, as
shown by the dashed histograms in Fig.~\ref{f:SP_sep}.

The energy range of the telescope divides naturally into the two
classes of events shown in Figs.~\ref{f:SP_dE/E}(a),(b). Protons
with kinetic energies between $2.3$ and $6.7\,$MeV traverse the
60$\,\mu$m surface-barrier detector and stop in the second layer
whereas those that stop in the final layer have energies up to
$31\,$MeV. The identification of protons by the $E$-$\Delta E$
method is, however, complicated in the transition region
between these two ranges. As shown in Fig.~\ref{f:SP_dE/E12}, the
band arising from deuterons traversing the first two layers, but
missing the third, overlaps with that from stopped protons. Taking
into account the finite energy resolution, protons are only
completely distinguished from deuterons in the low energy range of
Fig.~\ref{f:SP_dE/E12} for $T_p<4.4\,$MeV. On the other hand, the
lower limit that we have taken for the second range, $T_p>8\,$MeV,
has been set by the straggling considerations discussed below. The
overlap of the deuteron band imposes no restriction since
deuterons with energies above $\approx30\,$MeV are not scattered
into the acceptance of the system.\clearpage

The 60$\,\mu$m layer has no spatial resolution but the 300$\,\mu$m
and 5$\,$mm detectors each have strips on one of their sides.
These are placed perpendicular to the beam direction, leading to a
purely geometrical angular resolution $\sigma(\theta)$ of about
$0.4^\circ$ in the horizontal plane. Now, within our geometry, the
polar angle can be well approximated by this horizontal angle.
However, even for the relatively fast deuterons used to measure
the luminosity and relative alignment to the target, straggling of
more than 1$^\circ$ dominates the resolution. The angle of the
protons from the lower energy range, where the only position
information is provided by the strip number in the second layer,
is determined using the relative position of the target centre.
The finite target size then fixes the spectator polar angle to
within $\pm 5^\circ$. There remains an acceptable background of a
few per cent of events not originating from the target. For the
higher energy protons we have chosen a lower limit of 8$\,$MeV in
order to limit the straggling to be below 3$^\circ$.  The vertical
angle is fixed only by the target and detector sizes but this
corresponds essentially to the azimuthal angle, which has no
influence on the determination of the c.m.\ energy. The
characteristics of the two detection ranges are summarised in
Table~\ref{t:SP}.\clearpage

\begin{table}[H]
\caption{Effective ranges of detection, angular resolution and band separation
in terms of FWHM in the silicon telescope}
\vspace{3mm}
\begin{center}
\begin{tabular}{l|ll}
\hline
Combination  & $1^\mathrm{st}$ \& $2^\mathrm{nd}$\phantom{xxxxxx} &
$2^\mathrm{nd}$ \& $3^\mathrm{rd}$ \\
\hline
$T_p$-range      & 2.6 -- 4.4$\,$MeV & 8 -- 22$\,$MeV \\
$p_p$-range      & 70 -- 91$\,$MeV/c & 123 -- 204$\,$MeV/c \\
$\sigma(\theta_p)$ & $\approx 5^\circ$ & $\leq 3^\circ$ \\
Band separation & $> 2\, \times \,$FWHM & $>6\, \times \,$FWHM \\
\hline
\end{tabular}
\label{t:SP}
\end{center}
\end{table}

In order to study the experimental energy and angular resolution
of the silicon layers \textit{prior} to the measurements at COSY,
a dedicated experiment was performed using the proton beam at the
Tandem Accelerator of the Institute of Nuclear Physics of the
University of Cologne~\cite{DiplIL}. This led to the choice of a
suitable first layer detector. It also showed that the position in
the 5$\,$mm detector, read out using a resistor chain, could be
reconstructed with an accuracy better than 300$\,\mu$m in absolute value. The
angular resolution is then dominated by angular straggling, which
was measured to be in good agreement with results from Monte-Carlo
simulations~\cite{GEANT3}.

The target chamber of ANKE is placed just in front of the
spectrometer magnet D2 such that the low energy protons and
deuterons detected by the telescope are perturbed by the stray
magnetic field. Though the fields in this region, $B\lesssim
0.1\,$T, are much smaller than the maximum values in the
spectrometer, $B=1.6\,$T, significant corrections have to be
introduced because of the low spectator proton momenta.
Corrections of 0.6$^{\circ}$ to 1.8$^{\circ}$ relative to straight
lines, depending upon the spectator energy, have been
obtained from Monte Carlo simulations~\cite{GEANT3}.

\section{Measurement of the \boldmath{$pd\to p_\mathrm{sp}d \pi^0$}
Reaction}

\begin{figure}[H]
\begin{center}
\epsfig{file=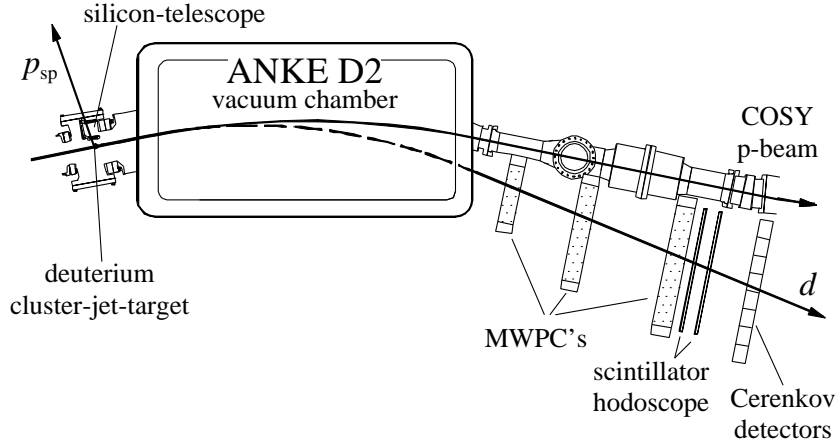,width=11cm} \caption{The part of the
ANKE set-up used for the measurement of the $pd\to p_\mathrm{sp}d
\pi^0$ reaction. Sketched are the circulating COSY proton beam,
the target chamber with the telescope for spectator proton
detection, the vacuum chamber of the spectrometer dipole magnet
D2, and the forward detection system used to track and identify
fast deuterons.} \label{f:anke}
\end{center}
\end{figure}

The magnetic dipole of ANKE, combined with the forward detector
system of multiwire proportional counters (MWPCs), scintillator
hodoscope and inclined \v{C}erenkov counters, allows us to
identify and measure the momenta of fast particles produced in
coincidence with spectator protons in the silicon telescope. This
system, illustrated in Fig.~\ref{f:anke}, is described in detail
in Ref.~\cite{ANKE}. The detection and identification of two
charged particles allows one to select the production of a single
meson, such as the $\pi^0$, $\eta$, or $\omega$, through the
missing mass technique.\clearpage

\begin{figure}[H]
 \begin{center} \epsfig{file=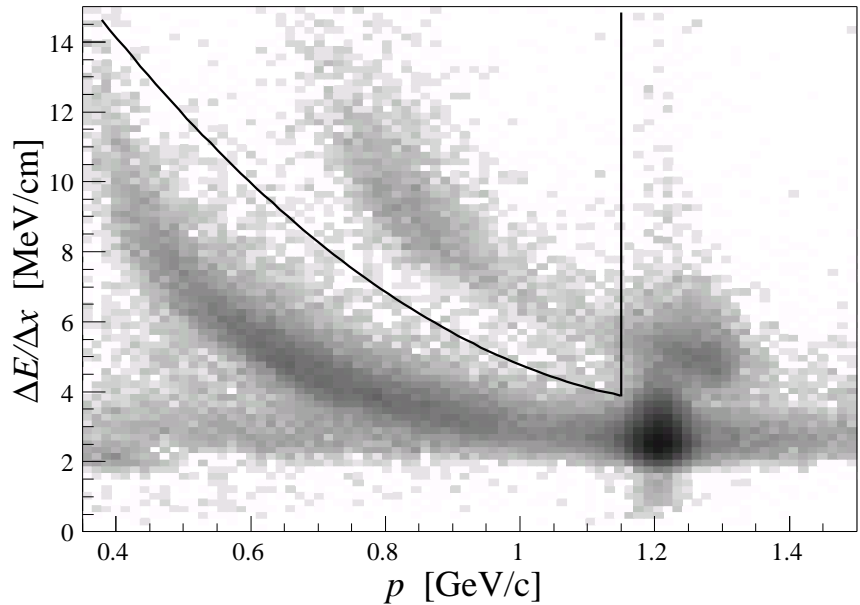,width=8cm,angle=270}

 \caption{The normalised energy loss per centimetre for particles in
 the first hodoscope layer of the ANKE forward detector {\em versus}
 their measured momentum. Clearly visible are the upper and lower
 bands originating from deuterons and protons respectively. Note that
 the statistics for protons are orders of magnitude greater than for
 deuterons. The lines indicate the cuts applied to select deuterons
 shown in Fig.~\ref{f:pi0_p}}

 \label{f:d_id}
 \end{center}
\end{figure}

To test this method of investigating meson production in
proton-neutron interactions at ANKE, we have
measured the $pd\to p_\mathrm{sp} d \pi^0$ reaction at a beam
momentum of 1.2$\,$GeV/c. In Fig.~\ref{f:d_id} the energy losses
of particles in the first plane of the scintillator hodoscope of
ANKE are plotted \textit{versus} their reconstructed momenta. The
spectrum is dominated by the proton peak around 1.2$\,$GeV/c
corresponding to small-angle deuteron break-up events. However,
one also observes clear proton and deuteron bands and, by imposing
a momentum-dependent threshold between them, one can reduce the
proton contribution significantly.  Since the Landau tail from the
quasi-elastic protons cannot be suppressed very effectively at
high momenta, only the range below 1.15$\,$GeV/c was selected when
extracting the $pn\to d\pi^0$ cross section, as indicated in
Fig.~\ref{f:d_id}.

\begin{figure}[H]
  \begin{center}
     \subfigure[Deuteron momentum distribution using the cuts indicated in
Fig.~\ref{f:d_id}]{
    \epsfig{file=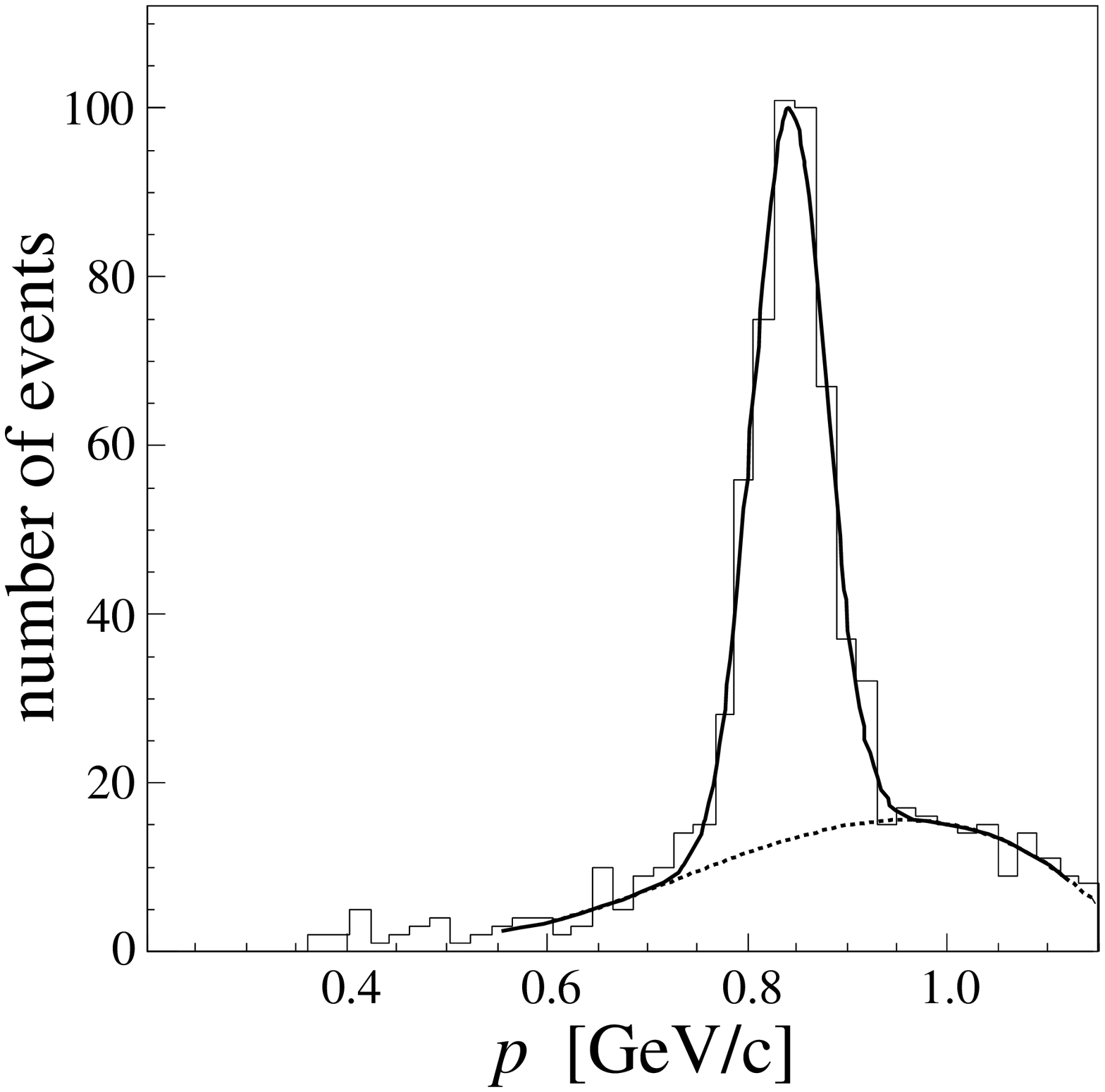,height=6.3cm}
    \label{f:pi0_p}}
     \subfigure[Missing mass distribution for the events in (a)]{
    \epsfig{file=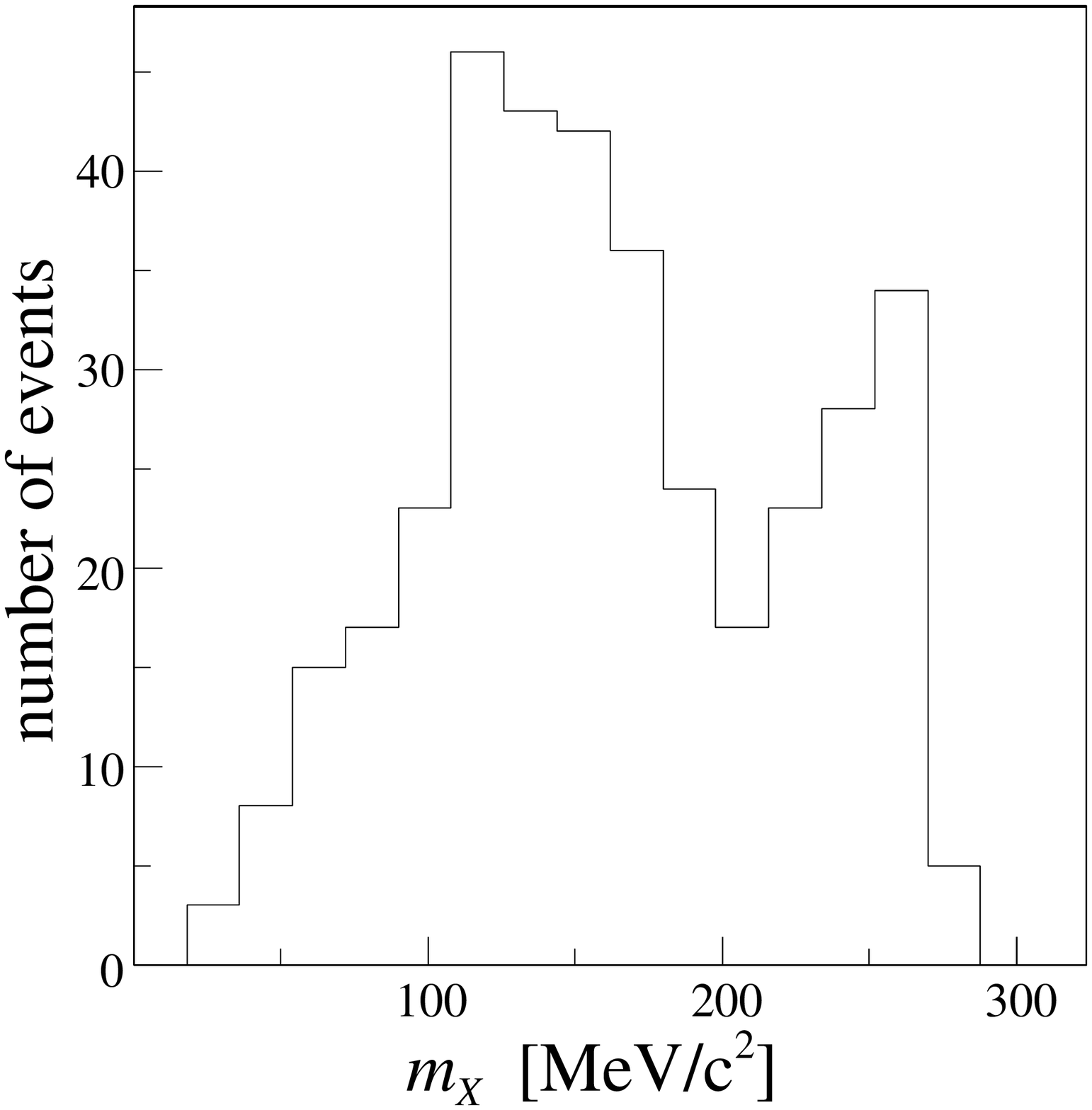,height=6.3cm}
    \label{f:pi0_mx}}

    \caption{Momentum and missing mass distributions using spectator
    protons with $2.6\leq T_\mathrm{sp} \leq 4.4\,$MeV and energy
    losses in the hodoscope for proton suppression.}

    \label{f:pi0}
  \end{center}
\end{figure}

The experimental momentum distribution is compared in
Fig.~\ref{f:pi0_p} to a Gaussian fit for the $d\pi^0$ events plus
a polynomial \textit{ansatz} for the background.  The number of
detected events for the $pn\to d\pi^0$ reaction has been deduced
solely from this figure where, unlike for the missing-mass
representation, the distribution is not biased by the assumption 
that the fast particle is a deuteron.  Since the experiment was
carried out very close to the two-pion threshold, the background
must arise primarily from protons misidentified as deuterons. This
is consistent with the missing mass distribution for the same
events, which is shown in Fig.~\ref{f:pi0_mx}.  There is a clear
peak at around 140$\,$MeV/c$^2$, corresponding to the undetected
$\pi^0$. The large width comes from measuring pion production well
above threshold with a limited angular resolution of the forward
array. The small shift compared to the true pion mass is
consistent with the geometric precision of the set-up. The
second peak towards the maximum missing mass is due to the
proton background.\\

The integrated luminosity for the $\pi^0$ production experiment at
600$\,$MeV was obtained by counting protons elastically and
quasi-elastically scattered from the target at laboratory angles
between 5$^{\circ}$ and 10$^{\circ}$, as described in
Ref.~\cite{Ko02}. It is not possible to separate cleanly $pd\to
pd$ from $pd\to ppn$ events using just the ANKE Forward Detector
information, though the momentum resolution is sufficient to
exclude pion production. The inclusive $pd$ differential cross
section can be reliably estimated within the Glauber
model~\cite{Roy}. Such a calculation, which takes into account the
sum of elastic and inelastic terms in closure approximation, could
be checked in part using the results obtained on the $pd\to pd$
differential cross section, which has been measured close to our
energy at $T_p=(582\pm 10)\,$MeV~\cite{pd_el}. It is believed that
the luminosity can be obtained to $\pm 7\%$ using this
procedure~\cite{Uzikov}.  The reduction in flux due to the
presence of a second nucleon in the target (shadowing), has been
taken into account by a 5\% correction, similar to that used for
$\eta$ production~\cite{Chia94}.

To calculate the acceptance of the whole system, the $pd\to
p_\mathrm{sp}d\pi^0$ reaction was simulated using the PLUTO event
generator~\cite{PLUTO}, which includes the Fermi motion in the
deuteron. The events thus obtained were fed into the standard
GEANT3 program~\cite{GEANT3}, adjusted to the geometry of the
ANKE. Finally, the momenta of both the spectator proton and fast
deuteron were subjected to the experimental cuts imposed by the
spectator telescope and forward detector system respectively.

After correcting for the tracking efficiency in the wire chambers
and the spectator detection, we deduce a total cross section of
$\sigma_\mathrm{tot}(pn\to d\pi^0)=(1.62\pm 0.14)\,$mb at an
effective mean beam energy of $T_\mathrm{beam} = 556\,$MeV. 
The c.m.\ energy is here reconstructd to 7$\,$MeV FWHM. A
direct measurement of this cross section with a neutron beam at
this energy gave $\sigma_\mathrm{tot}(np\to d\pi^0)=(1.6\pm
0.27)\,$mb~\cite{Wil71}. Alternatively, on the basis of isospin,
$\sigma_\mathrm{tot}(pn\to d\pi^0) =
\textstyle{\frac{1}{2}}\,\sigma_T(pp\to d\pi^+)= (1.53\pm
0.01)\,$mb, where we have used values from the standard data
compilation~\cite{SAID}. The agreement with our result is
therefore very satisfactory.

\section{Conclusions}

We have here described the construction and application of a solid
state telescope capable of measuring slow protons and deuterons in
the ultra-high vacuum conditions pertaining at an internal-target
station of a storage ring. The technique enables us not only to
identify the particles but, also to measure their kinetic energies
with 1\% precision. Using the tracking capabilities of the set-up
the angle of the spectator can be determined and the relative
positions of the target and all detectors measured, such that
geometrical uncertainties are minimised.

The combination of the silicon telescope with a magnetic analysis
system for fast particles permits the identification of reactions
on the neutron in the deuterium target and the measurement of
their absolute cross sections. The centre-of-mass energies in such
reactions are well defined by the kinematics of the spectator
proton and, in fact, the deuteron Fermi motion allows a scan over
a wide range of energies while keeping the beam momentum fixed.
Apart from the $pn\to d\pi^0$ test experiment reported here, the
system has already been used to investigate near-threshold
$\omega$~\cite{omega} and $\eta$~\cite{eta} production from the
neutron.

A further use of this technique is the monitoring of the
luminosity in experiments at storage rings. The energy of a recoil
deuteron detected near $90^\circ$ determines very well the
kinematics of elastic proton-deuteron scattering, thus fixing the
luminosity in terms of the $pd\to pd$ differential cross section.
This method was employed in the analysis of our $pn\to d\omega$
measurement~\cite{omega} and other reactions at ANKE. With the
current set-up it could not be used for the $pn\to d\pi^0$
measurement presented here. This geometric limitation can be
overcome by an optimised positioning of the system within the
redesigned vacuum chamber at ANKE.  It is important to note that
the telescope can also be used to evaluate the luminosity for $pp$
reactions with a hydrogen target by measuring the low energy
proton from $pp$ elastic scattering near $90^\circ$.

Another application of the set-up is the measurement of the beam
polarisation, which is becoming increasingly important in view of
the number of polarisation measurements proposed at
COSY~\cite{pol_prop}.  It was proven, using the telescope
described here, that it was possible to do this in parallel to
data-taking without any harm to or restriction on the
beam~\cite{pol_ann}.

The experience gained in building and using the spectator counter
system has been put to use in the design of second-generation
silicon telescopes~\cite{omega2}. Putting four of these new
telescopes even closer to the beam will increase the geometrical
acceptance by a factor 40.  Since all the detectors, especially
the 65$\,\mu$m layers, will be double-sided strip detectors, the
angles of even low energy spectators will be measured to
$\sigma(\theta) = \sigma(\phi) = 1^\circ-3^\circ$. This means that
the tracks from the extended target region of a polarised gas
target could also be reconstructed.  This will open the window to
use polarised deuterium as a source for polarised proton-polarised
neutron physics.

\section{Acknowledgements}

We would like to thank T.~Krings, G.~Fiori, and M.~Metz for designing,
engineering and testing the 5$\,$mm thick silicon detectors.
Moreover, we are grateful to W.~Borgs, G.~d'Orsaneo, P.~Wieder for
the support in the design of the support structure for our
detectors. H.~Hadamek and the mechanical workshop of the IKP build
this pieces. For the support in the measurements at the University
of Cologne, we thank A.~Dewald, L.~Steinert, and H.~Paetz
gen.~Schieck. Discussions with T.~Johansson on the CELSIUS
detectors were most helpful. The work has been financially
supported by the the FZ-J\"ulich (COSY-031, COSY-064).


\begin{thebibliography}{99}
\bibitem{Tord} R.~Bilger et al., Nucl.\ Instr.\ Meth.\ A~457
(2001) 64.
%
\bibitem{COSY11} P.~Moskal and T.~Johansson, COSY proposal 100 (2001)
(unpublished).
%
\bibitem{omega} S.~Barsov et al., Eur.\ Phys.\ J.\ (\textit{in press}),\\
\verb=http://de.arXiv.org/abs/nucl-ex/0305031=
%
\bibitem{PhDIL} I.~Lehmann, PhD thesis, University of Cologne, 2003.
%
\bibitem{eta} N.~Lang, IKP Annual Report 2003, Forschungszentrum J\"ulich; \\
A.~Khoukaz, COSY proposal 94 (2000) (unpublished),\\
\verb=http://www.fz-juelich.de/ikp/anke/en/proposals.shtml= .

\bibitem{ANKE} S.~Barsov et al., Nucl.\ Instr.\ Meth.\ A~462/3 (2001) 364.
%
\bibitem{Protic} G.~Riepe and D.~Proti\'{c}, Nucl.\ Instr.\ Meth.\
A~226 (1984) 103.
%
\bibitem{DiplIL} I.~Lehmann, Diploma thesis, University of Cologne (2000);\\
Internal Report, Forschungszentrum J\"ulich: FZJ-IKP-IB-E2-1/2000; \\
\verb=http://www.fz-juelich.de/ikp/anke/en/theses.shtml= .
\bibitem{res_chain} P.A.~Schlosser et al., IEEE Trans.\  Nucl.\ Sci.,
vol.\ 21, no.\ 1 (1974) 658.
%
\bibitem{SRIM} J.~F.~Ziegler, J.P.~Biersack, and U.~Littmark,
\textit{The Stopping and Range of Ions in Solids}, Pergamon Press
(N.Y., 1985); \verb=http://www.srim.org=
%
\bibitem{GEANT3} GEANT3, CERN Program Library W5013, CERN, 1993;\\
\verb=http://wwwinfo.cern.ch/asdoc/geant_html3/geantall.html=
%
\bibitem{Ko02} V.~Komarov et al., Phys.\ Lett.\ B~553 (2003) 179.
%
\bibitem{Roy} V.~Franco and R.~J.~Glauber, Phys.\ Rev.\ 142 (1966)
1195.
%
\bibitem{pd_el} E.~Boschitz et al., Phys.\ Rev.\ C~6 (1972) 457.
%
\bibitem{Uzikov} Yu.~Uzikov, IKP Annual Report 2001,
Forschungszentrum J\"ulich, p.~66; (private communication).
%
\bibitem{Chia94} E.~Chiavassa et al., Phys.\ Lett.\ B~337 (1994) 192.
%
\bibitem{PLUTO} PLUTO, A Monte Carlo simulation tool, GSI, 2000\\
\verb=http://www-hades.gsi.de/computing/pluto/html/PlutoIndex.html=
%
\bibitem{Wil71} S.~S.~Wilson et al., Nucl.\ Phys.\ B~33 (1971) 253.
%
\bibitem{SAID} R.~A.~Arndt et al., Phys.\ Rev.\ C~48 (1993) 1926;\\
\verb=http://gwdac.phys.gwu.edu/analysis/pd_analysis.html=
%
\bibitem{pol_prop} V.~Komarov, COSY proposal 20 (1999) (unpublished),\\
A.~Kacharava and F.~Rathmann, COSY proposal 125 (2003) (unpublished),\\
\verb=http://www.fz-juelich.de/ikp/anke/en/proposals.shtml= .
%
\bibitem{pol_ann} S.~Yaschenko et al., IKP Annual Report 2002,
Forschungszentrum J\"ulich.
%
\bibitem{omega2} R.~Schleichert et al., IEEE Trans.\  Nucl.\ Sci.,
vol.\ 50, no.\ 3 (2003) 301;\\
R.~Schleichert, COSY proposal 114 (2002) (unpublished),\\
\verb=http://www.fz-juelich.de/ikp/anke/en/proposals.shtml= .
%
\end{thebibliography}
\end{document}